\documentclass[prb,amsmath,amssymb,twocolumn,superscriptaddress,floatfix]{revtex4}
\usepackage{graphicx}
\usepackage{dcolumn}
\usepackage{bm}
\usepackage{subfigure}
\usepackage{amsmath}
\usepackage{feynmf}
\usepackage{hyperref}

\newcommand{\ua}{\uparrow}
\newcommand{\da}{\downarrow}
\newcommand{\ra}{\rightarrow}

\newcommand{\bs}{\boldsymbol}

\newcommand{\SRO}{Sr$_2$RuO$_4$}
\newcommand{\POS}{PrOs$_4$Sb$_{12}$}
\newcommand{\PPG}{PrPt$_4$Ge$_{12}$}
\newcommand{\bk}{\boldsymbol k}

\usepackage{times}

\begin{document}
\title{Spontaneous surface current in multi-component cubic superconductors with time-reversal symmetry breaking}
\author{Jia-Long Zhang}
\address{State Key Laboratory of Optoelectronic Materials and Technologies, School of Physics, Sun Yat-sen University, Guangzhou, 510275, China}
\author{Wen Huang}
\email{huangw001@mail.tsinghua.edu.cn}
\address{Institute for Advanced Study, Tsinghua University, Beijing, 100084, China}
\author{Dao-Xin Yao}
\email{yaodaox@mail.sysu.edu.cn}
\address{State Key Laboratory of Optoelectronic Materials and Technologies, School of Physics, Sun Yat-sen University, Guangzhou, 510275, China}
\date{\today}

\begin{abstract}
In this work we present a comprehensive study of the spontaneous currents in time-reversal symmetry breaking (TRSB) multi-component superconductors with cubic crystalline symmetry. We argue, not limiting to cubic lattices, that spontaneous current on certain high-symmetry surfaces can exist {\it only if} the TRSB pairing simultaneously breaks a certain pair of mirror symmetries. This is shown to be in exact correspondence with the Gingzburg-Landau (GL) theory and is verified by numerical Bogoliubov de-Gennes (BdG) calculations. In the course we extend the BdG to include effects of gap anisotropy and surface disorder, both of which could lead to much suppressed current. The GL theory has been known to describe well the spontaneous current. However, we highlight a special case where it becomes less adequate, and show that a refined effective theory for low temperatures is needed. These results could shed light on the phenomenology of cubic superconductors such as U$_{1-x}$Th$_x$Be$_{13}$, the filled skutterudites \POS, \PPG~and related compounds. 
\end{abstract}

\maketitle
\section{Introduction}
\label{sec:intro}
In cubic and tetrahedral superconductors, the peculiar crystalline symmetry allows for superconducting order parameters which belong to a plethora of multi-dimensional representations not accessible in other systems \cite{Sigrist:91,Mukherjee:06}. Some of these states may exhibit nontrivial topological properties and support exotic excitations such as protected surface modes and Majorana fermions \cite{Hasan:10,Qi:11}.  

In some cases, a time-reversal symmetry breaking (TRSB) multi-component pairing is stabilized deep in the superconducting state, such as in U$_{1-x}$Th$_x$Be$_{13}$ \cite{Smith:84,Ott:85,Ott:86,Heffner:86}, as well as the filled skutterudites \POS~\cite{Bauer:02,Aoki:03,Levenson:16}, \PPG~\cite{Gumeniuk:08,Maisuradze:10} and related compounds \cite{Shu:11,Zhang:15}. However, consensuses are still lacking regarding their exact pairing symmetries \cite{Pfleiderer:09}. Notably, these superconductors may generate spontaneous charge current at sample surfaces, domain walls separating regimes of distinct TRSB pairings and around crystalline defects, opening a unique perspective to peer into their exotic Cooper pairing. An effort along this line has indeed been made for \POS, which however did not find any definitive evidence of spontaneous surface current \cite{Hicks:10}. The null result resembles the situation \cite{Kirtley:07,Hicks:10,Curran:14} in the widely-studied putative chiral $p$-wave superconductor \SRO~\cite{Maeno:94}, which is also expected to support finite surface current \cite{Matsumoto:99}. While there have been a number of theoretical attempts to address this particular puzzle in \SRO, both within \cite{Ashby:09,Sauls:11,Imai1213,Bouhon:14,Lederer:14,Huang:15,Scaffidi:15,Etter:17} and outside \cite{Huang:14,Tada:15} the framework of chiral $p$-wave pairing, much less has been done for \POS.

There are indeed TRSB superconductors which does not support surface current at certain edges. One typical example is the $s+id_{x^2-y^2}$-wave superconductor at surfaces parallel to $x$ and $y$-axis (e.g. [100] surfaces) \cite{Furusaki:01}. However, it is known to exhibit finite spontaneous current around impurities and at the [110]-surface where it resembles an $s+i d_{xy}$ pairing \cite{Furusaki:01,Lee:09}. Furthermore, the $s+is$ state, which has been proposed for some iron-based superconductors, may generate current if fractional fluxes are pinned at domain walls between regions of $s+is$ and $s-is$ pairings \cite{Garaud:14,Garaud:16} or when the lattice rotational symmetry is further broken \cite{Maiti:15}, whilst no current may arise in an undistorted lattice \cite{Mahyari:14}. 

The primary objective of this work is to present a comprehensive study of the surface current in a multi-component TRSB superconductor with cubic symmetry. In particular, given the rich variety of superconducting phases available in these systems, it is tempting to ask whether any of their TRSB pairings would be free of spontaneous currents. For these purposes, we combine symmetry analyses, BdG calculations and a Gingzburg-Landau theory. The focus will be on the spontaneous currents on high-symmetry [100]- and [110]-surfaces. Since each of these two surfaces is invariant under at least one mirror reflection (orthogonal to the surface), and since their associated [100]- and [110]-planes are themselves mirror planes (see Fig. \ref{fig:sketch}), we name them mirror-invariant surfaces (MIS) for convenience. We will show that spontaneous current can emerge on these surfaces {\it only if} the TRSB pairing simultaneously breaks the two corresponding mirror symmetries. The argument is in fact of broad relevance to other systems with the required crystalline mirror symmetries, and is shown to be consistent with BdG calculations and GL analyses. In the course we also generalize the BdG calculations to include the effects of gap structure anisotropy and surface disorder. It is found that these two factors in general lead to a suppressed surface current, as is in line with the previous studies of chiral $p$-wave pairing \cite{Ashby:09,Lederer:14,Huang:15,Scaffidi:15}. More detailed investigation into the effect of surface disorder can be found in \onlinecite{Suzuki:16,Wang:18}.

The qualitative agreement between BdG and GL has been well recognized \cite{Lederer:14,Bouhon:14,Huang:14}. However, in this study we identify a special case where two phases described by almost equivalent GL theories turn out to produce markedly different surface currents in BdG calculations. As we shall see, this is due to the insufficiency of the GL description deep in the superconducting state, and an effective field theory more appropriate for low-$T$ readily accounts for the discrepancy. 

The paper is organized as follows. Sec. \ref{sec:symmetry} presents a symmetry analysis, where we show that the existence of spontaneous current on a MIS is dictated by the property of the superconducting pairing under two separate mirror reflections. Sec. \ref{sec:BdG} presents our tight-binding BdG calculations on a cubic lattice. We also present here calculations which take into account the effects of gap anisotropy and surface disorder. Sec. \ref{sec:GL} provides a general phenomenological GL description. Here we also highlight the case where GL becomes less adequate and present an alternative effective theory derived from a low-$T$ expansion. Finally, the results are briefly summarized in Sec. \ref{sec:discussions}.

\section{Symmetry analyses}
\label{sec:symmetry}
The most general form of the gap function of a multi-component superconducting state in a centrosymmetric system reads, 
\begin{eqnarray}
&&\hat{\Delta}_{\bs k} = \sum_i \phi_i h_i(\bs k)\cdot i \sigma_y \,, ~~~~~~~~~~~~~~~~~ \text{even-parity} \nonumber \\
&&\hat{\Delta}_{\bs k} = \sum_i \phi_i[\vec{d}_i(\bs k)\cdot \vec{\sigma}]\cdot i \sigma_y \,. ~~~~~~~~~~~ \text{odd-parity} 
\label{eq:GapFunction}
\end{eqnarray}
Here the $\phi_i$'s stand for the order parameter components and $h_{i}({\bs k})$ and $\vec{d}_{i}({\bs k})$ the respective even- and odd-parity basis functions belonging to certain irreducible point group representations. Usually, these basis functions form a single multi-dimensional representation, although mixed-representation pairing is also possible, such as the $s+d$ and $s+id$ pairings. It is worth pointing out that here we work in the band basis, where only intraband Cooper pairing is present in the weak coupling limit. This differs from the orbital-basis language adopted in some studies (e.g. Refs. \onlinecite{Zhou:08,Wan:09,Fischer:15,Bzdusek:17,Yanase:16,Wang:17}), although we will not dwell upon the distinctions. Also note that mixed-parity pairings (which violate inversion symmetry) are not considered here, but will be presented elsewhere. Listed in Table \ref{tb:irrep} are the basis functions of the irreducible representations of the cubic group $O_h$ \cite{Sigrist:91,Mukherjee:06}, which shall later become the focus of the present study. Time-reversal symmetry is broken if $\bs \phi=\{\phi_i\} \neq \bs \phi^\ast$. 

\begin{table}[h]
\centering
\caption{Irreducible representations and corresponding basis functions for even- and odd-parity states of a superconductor with $O_h$ symmetry \cite{Sigrist:91,Mukherjee:06}.}
\begin{tabular}{c c}
\hline\hline
Irrep & Basis function  \\
\hline
$A_{1g}$ & $k^2_x+k^2_y+k^2_z$  \\
$A_{2g}$ & $(k^2_x-k^2_y)(k^2_y-k^2_z)(k^2_z-k^2_x)$ \\
$E_{g}$  & $2k^2_z-k^2_x-k^2_y, k^2_x-k^2_y$ \\
$T_{1g}$ & $k_yk_z(k^2_y-k^2_z),k_zk_x(k^2_z-k^2_x),k_xk_y(k^2_x-k^2_y)$ \\
$T_{2g}$ & $k_yk_z,k_xk_z,k_xk_y$ \\
\hline
$A_{1u}$ & $k_x\hat{x}+k_y\hat{y}+k_z\hat{z}$  \\
$A_{2u}$ & $k_x(k^2_z-k^2_y)\hat{x}+k_y(k^2_x-k^2_z)\hat{y}+k_z(k^2_y-k^2_x)\hat{z}$ \\
$E_{u}$  & $2k_z\hat{z}-k_x\hat{x}-k_y\hat{y},k_x\hat{x}-k_y\hat{y}$ \\
$T_{1u}$ & $k_z\hat{y}-k_y\hat{z},k_x\hat{z}-k_z\hat{x},k_y\hat{x}-k_x\hat{y}$ \\
$T_{2u}$ & $k_z\hat{y}+k_y\hat{z},k_x\hat{z}+k_z\hat{x},k_y\hat{x}+k_x\hat{y}$ \\
\hline\hline
\end{tabular}
\label{tb:irrep}
\end{table}

\begin{figure}
\includegraphics[width=8cm]{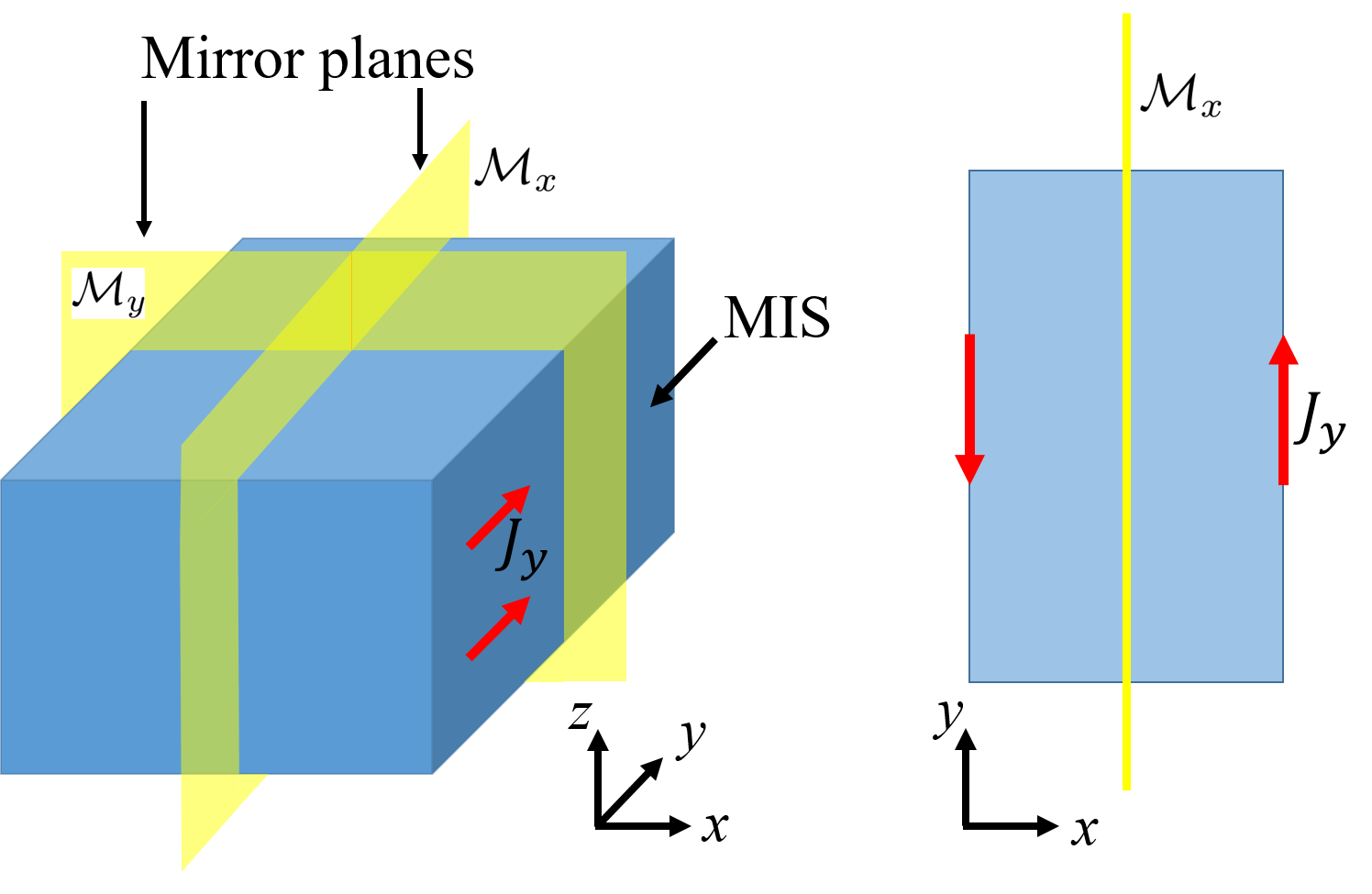}
\caption{(color online) Left: sketch of a mirror-invariant surface (MIS) defined in the main text, the associated pair of mirror planes (yellow) used to judge whether a particular component of the spontaneous surface current (red arrow) can exist. One of the two mirror planes is parallel to the surface ($\mathcal{M}_x$), while the other is perpendicular to the direction of the surface current component in question ($\mathcal{M}_y$). The blue block represents the superconductor under consideration, which is periodic in both $y$- and $z$-directions. Right: top view of the left panel. Note that, since we are considering pairings possessing inversion symmetry, and since the two opposite surfaces are related by inversion, the spontaneous current on the two surfaces, if any, shall flow in opposite directions. Surface current along $y$ (on the [100]-surface) is prohibited if the TRSB pairing preserves the reflection symmetry about either of the two mirror planes, i.e. $\mathcal{M}_x$ and $\mathcal{M}_y$ in the left panel. }
\label{fig:sketch}
\end{figure}

Through a simple symmetry argument, much can be learned about the existence of spontaneous currents on the MISs of a TRSB superconductor (assume the underlying crystal possesses the two mirror symmetries depicted in Fig. \ref{fig:sketch}). To set the stage for our discussions, we analyze a particular component of the current. The two relevant mirror operators can be denoted $\mathcal{M}_\perp$ and $\mathcal{M}_\parallel$, with the former (latter) describing a reflection perpendicular (parallel) to both the current and the surface in question. We shall show that this surface current component can arise only if the superconducting pairing simultaneously breaks the above stated mirror symmetries, i.e. 
\begin{eqnarray}
&&\mathcal{M}_{\perp}^{-1}\hat{\Delta}_{\bs k} \mathcal{M}_{\perp}^\ast \neq \pm \hat{\Delta}_{\bs k} \,,
\label{eq:sysCriterion}
\end{eqnarray}
and, 
\begin{eqnarray}
&&\mathcal{M}_{\parallel}^{-1}\hat{\Delta}_{\bs k} \mathcal{M}_{\parallel}^\ast \neq \pm \hat{\Delta}_{\bs k} \,.
\label{eq:sysCriterion1}
\end{eqnarray}

For concreteness, let's consider [100]-surface as a MIS and focus on the $y$-component of the spontaneous current, as sketched in Fig. \ref{fig:sketch}. In this case the relevant mirror operators $\mathcal{M}_\perp$ and $\mathcal{M}_\parallel$ become $\mathcal{M}_y$ and $\mathcal{M}_x$, whose associated mirror planes are $xz$ and $yz$, respectively. Formally, $\mathcal{M}_{\mu} = i \sigma_\mu \otimes R_\mu $, where $\sigma_\mu$ is the $\mu$-component Pauli matrix and $R_\mu$ denotes a reflection in spatial dimension perpendicular to the $\mu$-direction. Note that, in the full Nambu spinor basis $(c^\dagger_{\bs k,\ua},c^\dagger_{\bs k,\da},c_{-\bs k,\ua},c_{-\bs k,\da})$, the mirror operator takes the form,
\begin{equation}
\hat{\mathcal{M}}_\mu = \begin{pmatrix}
\mathcal{M}_\mu   & 0 \\
0 &   \mathcal{M}^\ast_{\mu}
\end{pmatrix} \,.
\end{equation}

To understand the condition set by Eq. (\ref{eq:sysCriterion}), assume
\begin{equation}
J^\text{tot}_y=  \langle \Omega |\hat{J}_y |\Omega\rangle \,,
\label{eq:curr1}
\end{equation}
where $\hat{J}_y \propto -i\partial_{y}$ is the $y$-component current operator and $|\Omega\rangle$ is the ground state wavefunction in the presence of an open boundary at the [100]-surface. Since the reflection reverses the direction of the current in $y$, i.e. $\{\hat{\mathcal{M}}_y, \hat{J}_y\}=0$, the mirror-reflected state $\hat{\mathcal{M}}_y|\Omega\rangle$ must satisfiy,
\begin{eqnarray}
J^\text{tot}_y  = -\langle \Omega|\hat{\mathcal{M}}_y^{-1} \hat{J}_y   \hat{\mathcal{M}}_y |\Omega\rangle  \,.
\label{eq:curr2}
\end{eqnarray}
If the superconducting pairing preserves the mirror symmetry, i.e. (up to an unimportant overall phase $\theta$) $\hat{\mathcal{M}}_y|\Omega\rangle = e^{i\theta}|\Omega\rangle$ or $\mathcal{M}_{y}^{-1}\hat{\Delta}_{\bs k} \mathcal{M}_{y}^\ast = \pm \hat{\Delta}_{\bs k}$, combining Eqns. (\ref{eq:curr1}) and (\ref{eq:curr2}) it follows that $J^\text{tot}_y$ must vanish. However, if $J^\text{tot}_y \neq 0$, the original and the mirror-reflected states must be distinct TRSB states -- a statement equivalent to Eq. (\ref{eq:sysCriterion}). 

In like manner, the breaking of the reflection symmetry about the mirror plane parallel to the surface, i.e. Eq. (\ref{eq:sysCriterion1}), is also necessary for the existence of spontaneous current. This is most easily seen in a set-up with two opposite MIS's on the right panel of Fig. \ref{fig:sketch}. In a state that carries finite $J^\text{tot}_y$ on one MIS, the opposite MIS must see an opposite current due to inversion symmetry. As a result, the current must reverse sign under a reflection $\mathcal{M}_x$ about the center of the geometry. This then suggests that the pairing must also break this mirror symmetry, henceforth Eq. (\ref{eq:sysCriterion1}). 

To summarize, in general the $\mu$-th component surface current on a MIS is prohibited if the pairing is unvaried under a reflection about either the mirror plane parallel to the surface, or about the mirror plane perpendicular to the $\mu$-axis (Fig. \ref{fig:sketch}, left panel). Equations (\ref{eq:sysCriterion}) and (\ref{eq:sysCriterion1}), and their variants, therefore constitute the complete symmetry criterion. We shall see an exact correspondence of these constraints in Sec. \ref{sec:GL}.

These arguments are of general relevance and can immediately be shown to apply to some simple cases such as chiral $p$-wave $(k_x+ik_y)\hat{z}$, $s+id_{xy}$ and $s+id_{x^2-y^2}$ on square lattices \cite{Matsumoto:99,Furusaki:01} and their generalizations to other 2D and 3D lattice models with two mirror planes. Both $\mathcal{M}_{x}$ and $\mathcal{M}_y$ turn the former two pairings into their time-reversal counterparts, hence a finite $J_y$ could in principle arise. However, the last one is invariant under these operations, therefore $J_y$ is forbidden. Furthermore, since all of these pairings preserve $\mathcal{M}_z$, no current shall arise on the [001]-surface.
 
Notably, in some special cases such as the non-$p$-wave chiral states and some fine-tuned anisotropic chiral $p$-wave states, the two mirror symmetries in question are broken, yet the surface current may still vanish \cite{Huang:14,Tada:15,Huang:15}. Nevertheless, the vanishing in these cases are not protected. For example, the chiral $d$-wave pairing on a trigonal lattice can support finite surface current \cite{Huang:14}. 

The symmetry criterion also permits some affirmative statements about the spontaneous currents in TRSB cubic superconductors. We first list some representative TRSB states in these systems. The two-dimensional representations $E_g$ and $E_u$ permit states with ${\bs \phi} = \Delta_0(1,\pm i)$, while a three-dimensional $T_g$ or $T_u$ phase can take either $\bs \phi = \Delta_0(1,\pm i, 0)$ or $\bs \phi = \Delta_0(1,w, w^2)$ with $w=\pm 2i\pi/3$, as well as their equivalences. Additionally, there are mixed-representation states.

According to Eqns (\ref{eq:sysCriterion}) and (\ref{eq:sysCriterion1}), on the [100]-surface, the TRSB states in the $T_g$ and $T_u$ representations can support spontaneous surface current, while the $E_g$ and $E_u$ states cannot. However, on the [110] surface the two latter states can be expressed in a rotated frame, e.g. the TRSB $E_g$ state becomes $2k_z^2-k_x^2-k_y^2 \pm i k_x k_y$, which satisfies the criterion. Hence the $E_g$ and $E_u$ states can generate finite in-plane current on [110], although the $z$-component current still vanishes. It is also easy to check that for any mixed-representation TRSB state in the cubic group, there always exist MISs where spontaneous currents may arise. All of these, including others not enumerated here, can be verified in the numerical BdG calculations to be introduced in Sec. \ref{sec:BdG}.

Note that since there exists no mirror plane parallel to the [111]-surface, the above argument does not directly apply to this surface. However, the GL analyses in Sec. \ref{sec:GL} shall show that this surface can support finite current. 

Along similar lines, symmetry arguments also apply to the spontaneous currents around crystalline defects. In essence, spontaneous current may arise around the defects in a TRSB superconductor provided that the pairing breaks some discrete point group symmetries of the underlying lattice (rotation, mirror reflection, etc). We will not elaborate, but would only refer to the application in some previous case-by-case studies \cite{Lee:09,Maiti:15}.

\section{Bogoliubov de-Gennes calculations}
\label{sec:BdG}
We consider for simplicity a cubic lattice model with only nearest neighbor hopping $t$ whose dispersion takes the form $\xi_{\bk}=-2 t (\cos k_x+\cos k_y+\cos k_z)-\mu$, where $\mu$ sets the chemical potential. We numerically solve the BdG equations on a $N\times N\times N$ lattice for different superconducting pairings, with open boundaries in the $x$-direction ($x=0$ and $x=N$) and periodic boundaries in the other two. The gap functions $\hat{\Delta}_{\bk}$ in (\ref{eq:GapFunction}) assume the lattice-generalized forms of the basis functions given in Table \ref{tb:irrep}. 

In actual calculations, Fourier transformations along $y$ and $z$ are performed. The surface current is defined as,
\begin{equation}
\hat{j}_\mu(i)=-\frac{i t}{N^2} \sum_{\substack{\bs k_\parallel \\ \sigma = \ua,\da}}\left[ c^\dagger_{\bs k_\parallel,\sigma}(i) c_{\bs k_\parallel,\sigma}(i)-\text{H.c.}\right]\sin k_\mu \,,
\end{equation}
where $\mu=y,z$ and $\bs k_\parallel=(k_y,k_z)$ denotes the momentum parallel to the surface. Note that current is formally expressed in units of $et/\hbar$, but we have set $e/\hbar$ to unity. The total $\mu$-component surface current follows as,
\begin{equation}
\hat{J}^\text{tot}_\mu=\sum_{i=1,\frac{N}{2}}\hat{j}_\mu(i).
\end{equation}
In the following, we shall present the results for some representative pairing states.


\subsection{Total surface current}

We first study the surface current of various TRSB states at ideal surfaces. As the main purpose of the present subsection is to verify the conclusions obtained in Sec. \ref{sec:symmetry}, we shall take here the simplest lattice generalization of the pairing basis functions in Table \ref{tb:irrep}. For example, the gap component $k_x \hat{x}$ is replaced by $\sin k_x \hat{x}$, $k_x^2 -k_y^2$ by $\cos k_x - \cos k_y$, etc. The calculations shown for the [110]-surface assume gap functions expressed in a rotated coordinate basis, such as $(2\cos k_z-\cos k_x-\cos k_y, 2 \sin k_x \sin k_y)$ for $E_g$, and $(2 \sin k_z \hat{z}-\sin k_x \hat{x}- \sin k_y \hat{y}, \sin k_y \hat{x}+\sin k_x \hat{y})$ for $E_u$. 

\begin{figure}
\centering
\includegraphics[width=8cm]{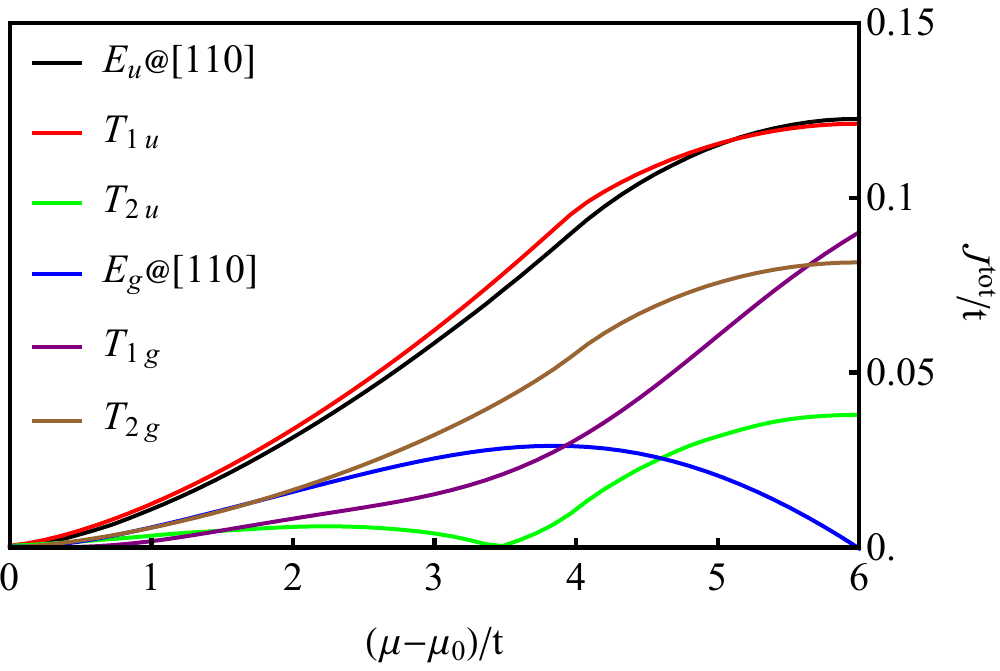}
\caption{(color online) Total surface current of various TRSB superconducting states as a function of chemical potential in a tight-binding BdG calculation. The $x$-axis is the chemical potential measured w.r.t. the band bottom $\mu_0=-6t$. For the two-dimensional representations, we have chosen ${\bs \phi} = \Delta_0(1, i)$, and for the three-dimensional representations, we select only those states with $\bs \phi = \Delta_0(1,w, w^2)$, where $w=\pm i2\pi/3$ and $\Delta_0=0.2t$. All calculations were performed at $T=0$. The calculations with effective [110]-surface are indicated.}
\label{fig:Jtot}
\end{figure}

The main results for a selected few states in the multi-dimensional representations are shown in Fig. \ref{fig:Jtot}, from where it is straightforward to deduce the complete agreement with the symmetry analyses in the previous section. Note that results are not plotted for the scenarios where the current vanishes, such as on the [100]- and [001]-surfaces of the $E_g$ and $E_u$ states. 

\subsection{Effects of gap anisotropy and surface disorder}
As has been emphasized in previous studies \cite{Ashby:09,Lederer:14,Huang:15,Scaffidi:15}, superconducting gap anisotropy and surface disorder could help explain the curious absence (or smallness) of the surface current in the putative chiral $p$-wave \SRO. Here we examine their effects on the surface current of a cubic TRSB superconductor. 

\begin{figure}
\centering
\includegraphics[width=8cm]{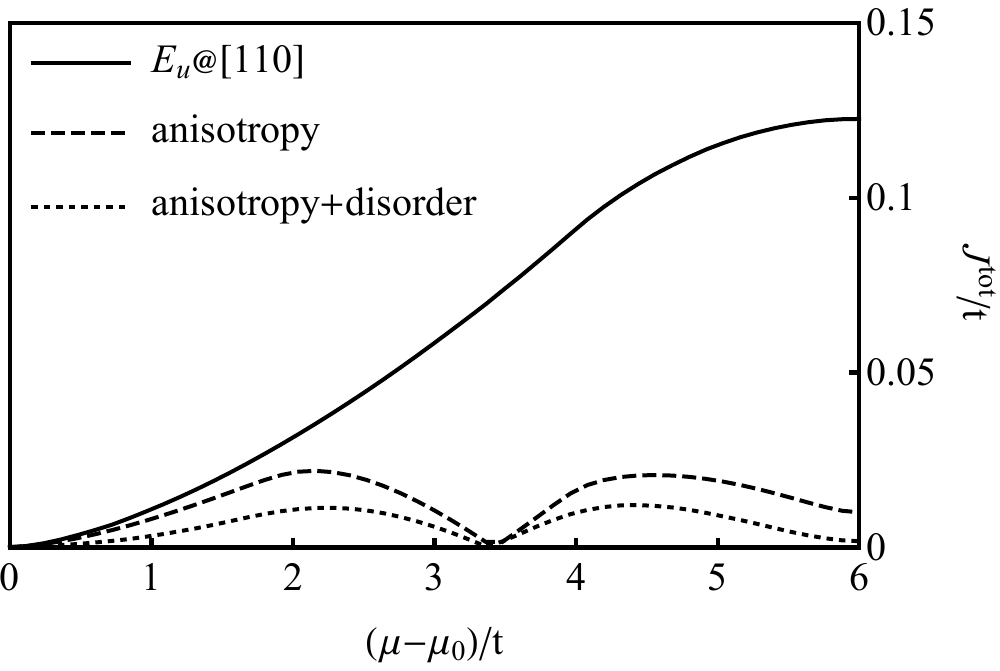}
\caption{Effect of gap anisotropy and surface disorder on the magnitude of the spontaneous current of the $E_u$ $(1,i)$ states at the [110]-surface. The simple and anisotropic $E_u$ pairing gap functions are both described in the text. Surface disorder is simulated by setting the pairing to be zero in the surface regime between the sites $i = 1$ and $10$. The calculations were carried out using the same parameters as in Fig. \ref{fig:Jtot}, except that the temperature here is $T=0.1\Delta_0$. Note with the anisotropic pairing the total current changes sign at around $\mu-\mu_0=3.4t$.}
\label{fig:currAni}
\end{figure}

The gap anisotropy can be modeled by generalizing to higher order lattice harmonics. For example, the $E_u$ pairing with $(2 \sin 2k_z \hat{z}-\sin 2k_x \hat{x}- \sin 2k_y \hat{y}, \sin 2k_y \hat{x}+\sin 2k_x \hat{y})$ shall exhibit stronger degree of gap anisotropy, in particular away from low fillings, compared to the simple gap function employed in the previous subsection. On the other hand, surface disorder can be implemented by setting the amplitude of the gap to be zero near the surface. Fig. \ref{fig:currAni} shows the results of a set of representative calculations of the $E_u$ state with an effective [110]-surface. As anticipated, both gap anisotropy and surface disorder lead to substantially suppressed spontaneous current. Similar effects can be shown to hold for other TRSB pairings. It is therefore tempting to attribute the null results \cite{Hicks:10} on \POS~to these two factors. Nevertheless, a sharp and disorder-free surface should still see finite spontaneous current, except in rare cases with fine-tuned pairing functions. 

\section{Ginzburg-Landau theory}
\label{sec:GL}

\subsection{General theory}
Multiple GL analyses of the surface current problem, mostly in the contexts of chiral and $s+id$ superconductors in two spatial dimensions, have been done in previous studies \cite{Ashby:09,Lederer:14,Bouhon:14,Huang:14,Etter:17}. Overall, the GL theory and the semiclassical BdG approaches have thus far reached excellent qualitative agreement. On this basis, in what follows we shall first briefly discuss the consistency between GL and the results in the preceding sections. After that we proceed to an interesting special case where the predictive power of GL becomes less affirmative and where a refined low-$T$ effective theory becomes necessary. We again concentrate on high-symmetry planes, taken to be the $yz$-plane to make contact with the preceding discussions. 
 
The generic form of the GL free energy, up to the quartic order, reads,  
\begin{equation}
f= f_0 + f_\nabla  \,,
\end{equation}
with the uniform free energy density,
\begin{eqnarray}
f_0 &=& \alpha_i |\phi_i|^2 + \beta_i |\phi_i|^4 + \beta_{ij}|\phi_i|^2|\phi_j|^2 \nonumber \\
&&+ \beta_{ij}^\prime(\phi_i^\ast\phi_j +\phi_j^\ast\phi_i)^2 + ...\,,
\label{eq:f0}
\end{eqnarray}
in which we have dropped terms such as $|\phi_i|^2\phi_i^\ast\phi_j$ disallowed by symmetry, and the gradient energies,
\begin{eqnarray}
f_\nabla&=&  k_{i\mu}^{j\nu} (\partial_\mu \phi_i)^\ast (\partial_\nu \phi_j) \,,
\label{eq:TuGLgradient}
\end{eqnarray}
where $i=1, ... ,N$ are the indices of the $N$ order parameter components and $\mu=x,y,z$ stand for the spatial coordinates. For multi-dimensional representations $\alpha_i \equiv \alpha \propto T-T_c$. Note that summation over the indices is left implicit for clarity. We shall continue to use this convention unless otherwise specified. 

The $k$-coefficients bear special significance in the following discussions. In the weak-coupling limit, for even-parity pairings, they are given by \cite{Huang:14,Huang:15}
\begin{eqnarray}
  k_{i\mu}^{j\nu} &=& k_{i\nu}^{j\mu} \nonumber \\
&=& \frac{T}{V}\sum_{w_n,\bs k} h_{i}(\bs k)h_j(\bs k) \times \nonumber \\
&& \frac{\partial^2}{\partial_{q_\mu}\partial_{q_\nu}}\left[ g(w_n,\bs k+ \frac{\bs q}{2}) \bar{g}(w_n,\bs k-\frac{\bs q}{2}) \right]_{\bs q \ra 0} \nonumber \\
&\propto & \left\langle v_{\mu,\bs k} v_{\nu,\bs k} h_{i}(\bs k)  h_{j}(\bs k)\right\rangle  \,,
\label{eq:kEven}
\end{eqnarray}
and analogously for odd-parity states, 
\begin{equation}
 k_{i\mu}^{j\nu} =  k_{i\nu}^{j\mu}  \propto  \left\langle v_{\mu,\bs k} v_{\nu,\bs k} [\vec{d}_{i}(\bs k)\cdot \vec{d}_{j}(\bs k)] \right\rangle  \,,
\label{eq:kOdd}
\end{equation}
where $g(w_n, \bs k) = (iw_n - \xi_{\bs k})^{-1}$ and $\bar{g}(w_n,\bs k)= (iw_n + \xi_{-\bs k})^{-1}$ with the Matsubara frequency $w_n = (2n+1)\pi T$, $T$ and $V$ denote respectively the temperature and volume of the system, $v_{\mu,\bs k}= \partial_{\bs k}\xi_{\bs k}$ the electron velocity, and $\langle ... \rangle$ stands for an average over the Fermi surface (same below). It is easy to verify that $k_{i\mu}^{i\nu} =0$ for $\mu \neq \nu$. For other surfaces, analogous expressions can be obtained using properly rotated frames, as mentioned in Sec \ref{sec:symmetry}. 

Similar to what has been discussed extensively in previous studies, the current density is related to the spatial modulations of the out-of-phase order parameter components as follows \cite{Ashby:09,Bouhon:14,Huang:14,Huang:15,Etter:17}, 
\begin{eqnarray}
j_{y} &=&\sum_{i,j} k_{ix}^{jy} \cdot \text{Im}[ (\partial_x \phi_i^\ast)\phi_{j}- \phi_i^\ast \partial_x \phi_{j} ] \,, ~~~(i\neq j) \nonumber \\
j_{z} &=& \sum_{i,j} k_{ix}^{jz} \cdot \text{Im}[ (\partial_x \phi_i^\ast)\phi_{j}- \phi_i^\ast \partial_x \phi_{j} ]\,, ~~~(i\neq j) 
\label{eq:GLcurrent}
\end{eqnarray}
Since different components generically exhibit distinct spatial modulations near the boundary, the existence of surface current is overwhelmingly dictated by the coefficients $k^{jy(z)}_{ix}$. It is now straightforward to crosscheck the conclusions in the previous sections. For example, if both $h_{i}(\bs k)$ and $h_{j}(\bs k)$ $[\text{or}~\vec{d}_i(\bs k)~ \text{and}~\vec{d}_j(\bs k)]$ are invariant or if they both change sign under mirror reflections $\mathcal{M}_x$ or $\mathcal{M}_y$, then $k_{ix}^{jy}$ and its corresponding contribution in (\ref{eq:GLcurrent}) must vanish \cite{footnote3}. In other words, the current is prohibited if the complex pairing respects the mirror symmetry about either $\mathcal{M}_x$ or $\mathcal{M}_y$. Finally, higher order terms in the free energy, such as $\partial_x^3\phi_i \partial_y \phi_j^\ast$, may be considered \cite{footnote2}. However, these contributions can also be shown to be dictated by the same mirror symmetries. In this respect, the GL theory reaches an excellent agreement with Secs. \ref{sec:symmetry} and \ref{sec:BdG}. 

As an important remark, the applicability of GL goes beyond the restrictive cases with two crystalline mirror planes. For example, the $E_g$ and $E_u$ states can be shown to have finite $k^{jy(z)}_{ix}$ for an effective [111]-surface, suggesting finite current on this surface. Generalizing GL to other non-high-symmetry surfaces, $k^{jy(z)}_{ix}$ and their variants in general do not vanish. It is therefore tempting to conjecture that spontaneous surface current should generically appear on non-high-symmetry and irregular surfaces of a TRSB superconductor, although we cannot give a rigorous proof. Further, the theory has also been applied to study the spontaneous current around point defects \cite{Lee:09,Maiti:15}.

\subsection{$T_{1u}$ Vs $T_{2u}$: partial failure of the GL theory}
Despite the overall satisfactory description by the GL theory, there are some interesting rare cases where it slips. In the present study, GL predicts that the $T_{1u}$ and $T_{2u}$ phases with $\bs \phi \sim (1,w,w^2)$ [and similarly $(1,i,0)$] should carry the same surface current (up to a sign difference) at the [100]-surface. To see this explicitly, first note that the two $T_u$ phases are characterized by exactly the same $\beta$-coefficients in (\ref{eq:f0}). Quoting Ref. \onlinecite{Huang:18},
\begin{eqnarray}
&& \beta_i \propto \left\langle |\vec{d}_{i}(\bs k)|^4 \right\rangle \,,  \\
&& \beta_{ij} \propto  \left\langle |\vec{d}_{i}(\bs k)|^2 |\vec{d}_{j}(\bs k)|^2 \right\rangle  \times 2  \,,\\
&& \beta^\prime_{ij} \propto  \left\langle [\vec{d}_{i}(\bs k)\cdot\vec{d}_{j}(\bs k)]^2 - |\vec{d}_{i}(\bs k) \times \vec{d}_{j}(\bs k) |^2 \right\rangle  \,,
\label{eq:betaOdd}
\end{eqnarray}
which can be shown to be the same for both $T_{1u}$ and $T_{2u}$. Their difference originates only from the gradient terms. The following relations hold, 
\begin{eqnarray}
&& k_{1x}^{1x}=k_{2y}^{2y}=k_{3z}^{3z} \propto \left\langle k_x^2(k_y^2+k_z^2)\right\rangle  \,,
\label{eq:TuK1} \\
&& k_{1y}^{1y}=k_{1z}^{1z}=k_{2x}^{2x}=k_{2z}^{2z} = k_{3x}^{3x}=k_{3y}^{3y}  \propto \left\langle k_y^2(k_y^2+ k_z^2) \right\rangle  \,,
\label{eq:TuK2} \nonumber \\
&&  \\
&& k_{1x}^{2y}=k_{1y}^{2x} = k_{2y}^{3z}=k_{3z}^{2y}= k_{3z}^{1x}=k_{1x}^{3z}\propto \pm \left\langle k_x^2k_y^2 \right\rangle \,,
\label{eq:TuKcross} \\
&& k_{i\mu}^{j\nu}=0 \,,~~~~~~~~\hfill{ \text{all others,}}
\end{eqnarray}
where in (\ref{eq:TuKcross}) ``$+$'' and ``$-$'' are taken for the $T_{1u}$ and $T_{2u}$, respectively. Up to this order, the cross-gradient terms associated with (\ref{eq:TuKcross}) are the only terms that distinguish the $T_{1u}$ and $T_{2u}$ phases. At the ideal [100] surface as in the BdG calculations, the spatial modulation of the order parameter components is governed by the joint action of $f_0$ and the gradient energies associated with (\ref{eq:TuK1}) and (\ref{eq:TuK2}). The cross-gradient terms associated with (\ref{eq:TuKcross}) have no impact in this matter due to the translational invariance parallel to the surface. It then follows that the two phases must observe the same spatially varying order parameters. Accordingly, the surface current of the two phases, for example the $y$-component $j_{y} = 2k_{1x}^{2y} \cdot \text{Im}[ (\partial_x \phi_1^\ast)\phi_{2}- \phi_1^\ast \partial_x \phi_{2} ] $, must differ only by a sign. 

\begin{figure}
\centering
\includegraphics[width=8cm]{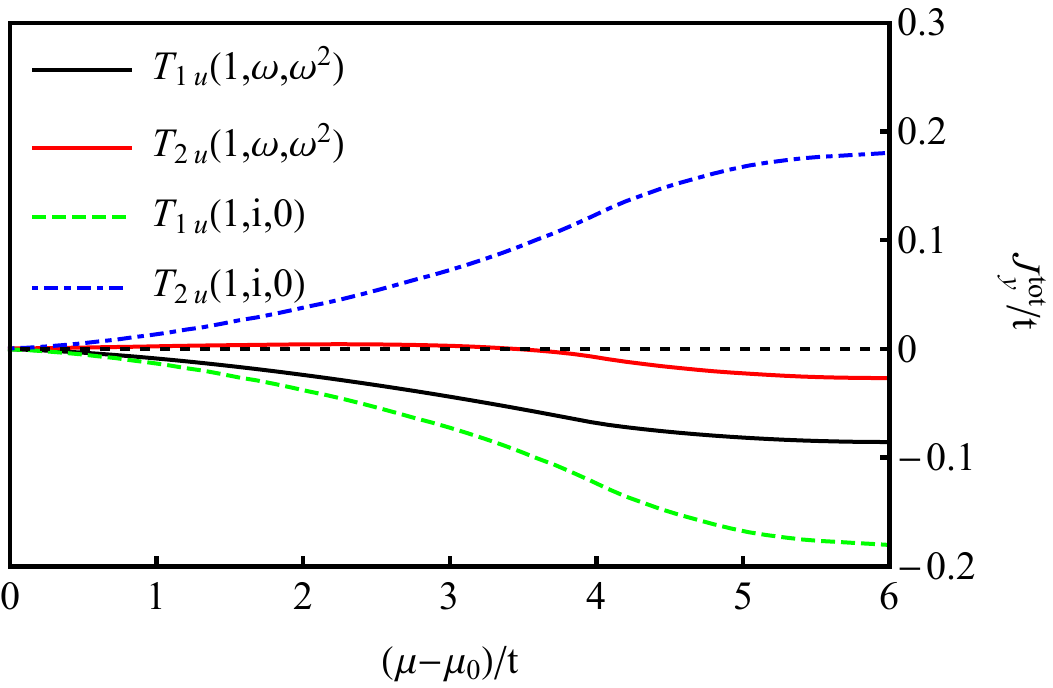}
\caption{BdG results: $y$-component of the total [100]-surface current of $(1,w,w^2)$ and $(1,i,0)$ states in the $T_{1u}$ and $T_{2u}$ representations. The gap functions and parameters used here are the same as in Fig. \ref{fig:Jtot}.}
\label{fig:JyTu}
\end{figure}

\begin{figure}
\includegraphics[width=6cm]{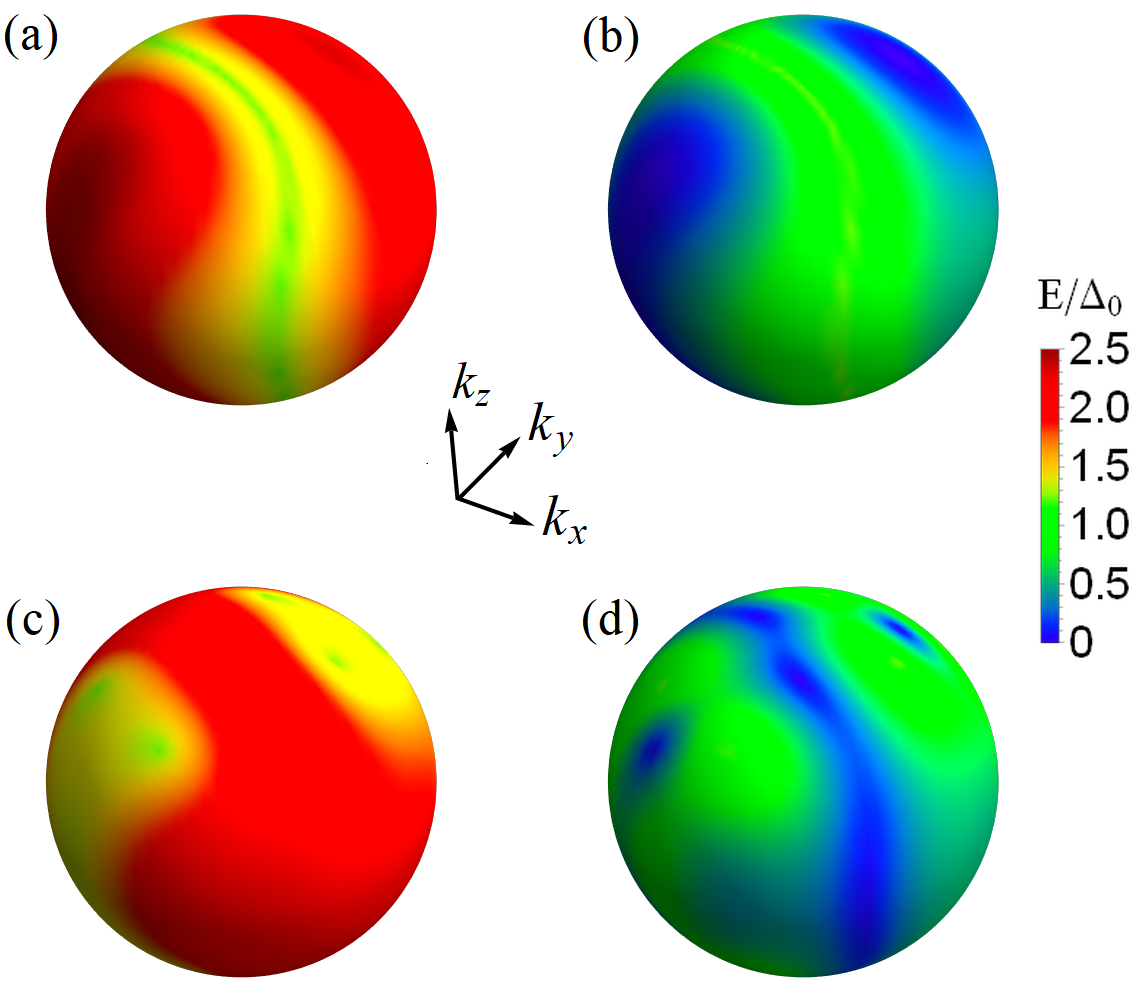}  
\caption{(color online) Contours of the gap structure of the $(1,w,w^2)$ states in the (a,b) $T_{1u}$ and (c,d) $T_{2u}$ representations. The calculations assume gap functions given by the simple bases as in Table \ref{tb:irrep} and a spherical Fermi surface. Two gaps appear for each of the non-unitary states.}
\label{fig:TuGaps}
\end{figure}

\begin{figure}
\includegraphics[width=8cm]{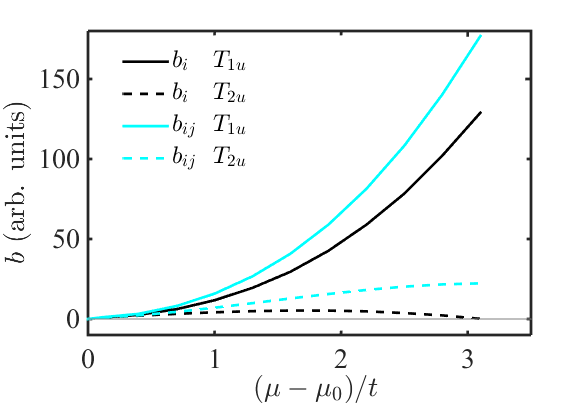}
\caption{(color online) Coefficients $b_i$ and $b_{ij}$ of the free energy (\ref{eq:T0F}) at low fillings of the lattice models (near the continuum limit) as in the BdG calculations. The $x$-axis is the chemical potential measured w.r.t. the band bottom. Calculation is performed at a somewhat elevated temperature $T=\Delta_0/5$ with $\Delta_0=0.4t$ for better convergence.}
\label{fig:F0GLcoefficients}
\end{figure}

This is in agreement with BdG for the two-component $T_u$ states $\bs \phi = \Delta_0(1,i,0)$, as in Fig. \ref{fig:JyTu}. However, for the three-component $(1,w,w^2)$ states, the edge currents of the $T_{1u}$ and $T_{2u}$ representations are markedly different (Fig. \ref{fig:JyTu}). This quantitative discrepancy can be attributed to the deficiency of the GL theory at low-$T$. Being a perturbative expansion in powers of the order parameters, the GL free energy is only exactly valid in the limit $|\phi_i| \ra 0$ near $T_c$ and is thus oblivious to the distinct gap structure and quasiparticle dispersion \cite{footnote1} in $T_{1u}$ and $T_{2u}$ as shown in Fig \ref{fig:TuGaps} (although GL does adequately capture the symmetry of the order parameters). They can be accounted for in an effective field theory appropriate for low-$T$, which can be obtained via an expansion in powers of the small deviations from the low-$T$ order parameters,
\begin{equation}
\phi_i \ra \phi_{i,0} + \varphi_i
\end{equation}
where the $\phi_{i,0}$'s and $\varphi_i$'s represent the mean-field bulk order parameter components and their fluctuations, respectively. It suffices to consider the uniform free energy density,

\begin{eqnarray}
F_0 &=& a_{i}|\varphi_i|^2 + \bar{a}_{i} \left[ (\varphi_{i})^2+(\varphi_{i}^\ast)^2 \right]   \nonumber \\
&& + b_i |\varphi_i|^4+ b_{ij} |\varphi_i|^2|\varphi_j|^2 +b^\prime_{ij}(\varphi_i^\ast \varphi_j + \varphi_j^\ast \varphi_i)^2  \nonumber \\
&& + \bar{b}_{ij} \left[ (\varphi_i \varphi_j)^2 + (\varphi_i^\ast \varphi_j^\ast)^2 \right]  \nonumber \\
&& + \bar{b}_{ij}^\prime \left[(\varphi_i)^2|\varphi_j|^2+ (\varphi_i^\ast)^2|\varphi_j|^2 \right] + ...\,, (i\neq j)
\label{eq:T0F}
\end{eqnarray}
Notice that since the expansion is performed with respect to a particular symmetry-broken state $(\phi_{1,0},\phi_{2,0},\phi_{3,0})= \Delta_0(1,w,w^2)$ in the bulk, the $U(1)$ symmetry is not preserved for the fields $\varphi_i$. Hence terms like those associated with $\bar{a}$, $\bar{b}$ and $\bar{b}^\prime$ are in general allowed. The dichotomy between the two $T_{u}$ phases is readily seen by noting their disparate coefficients in (\ref{eq:T0F}), as demonstrated in Fig. \ref{fig:F0GLcoefficients}. This naturally implies the different behavior of the $\varphi_i$ fields, henceforth different surface current in the $T_{1u}$ and $T_{2u}$ phases.

\section{Discussions and Summary} 
\label{sec:discussions}
In this paper we have given a consistent description of the spontaneous surface currents on the MISs of multicomponent TRSB superconductors with cubic symmetry. We showed that the surface current can arise only when the TRSB pairing simultaneously breaks the symmetry about a pair of mirror planes, one perpendicular and the other parallel to the surface, as summarized in Eqns (\ref{eq:sysCriterion}) and (\ref{eq:sysCriterion1}). The conclusion also applies to other TRSB superconductors possessing the relevant crystalline mirror symmetries. Based on the analyses, we also conjecture that the surface current is generally nonvanishing around crystalline defects, on non-mirror-invariant, non-high-symmetry, or irregular surfaces in generic models of multicomponent TRSB pairing with inversion symmetry. Further, we did not explore the spontaneous currents at the domain walls between regions of different TRSB pairings. Since such topological defects have been shown to induce finite spontaneous flux even in the simplest case of $s+is$ superconductors \cite{Garaud:14}, it is reasonable to expect spontaneous current at the domain walls of the more complicate TRSB phases discussed here. 

Throughout the work we mainly focused on the cubic $O_h$ group. This is appropriate for U$_{1-x}$Th$_x$Be$_{13}$. \POS~and \PPG~are characterized by the group $T_h$, which is a subgroup of $O_h$. Hence the number of possible multicomponent pairings in these two compounds is reduced. Nonetheless, the same analyses carry through. 

Note that we did not dwell upon the debate about the exact nature of the multicomponent pairing in these compounds \cite{Pfleiderer:09}, which goes beyond the scope of the present study. Irrespective of this, our study suggests that, gap anisotropy combined with possible surface disorder may hold the key to explain the null results on the surface current in the scanning SQUID measurements of \POS~\cite{Hicks:10}. Noteworthily, here we have ignored the possible multi-band character of the system, under which circumstance the spontaneous current may also be drastically influenced by interband interferences at the surface \cite{Zhang:17}.  

Finally, while the qualitative power of the GL theory is unquestionable, we identified a special case where two TRSB states, predicted by GL to carry the same surface current, instead yield drastically different outcome in low-$T$ BdG calculations. This calls for caution when using GL to infer some low-$T$ properties. We showed that the quantitative discrepancy originates from the oblivion of the quasiparticle gap structure deep in the superconducting state, and that a low-$T$ expansion is needed for more accurate descriptions. 

\acknowledgements
We would like to thank Egor Babaev, Shaokai Jian, Shuai Yin, and in particular Tom\'a\v{s} Bzdu\v{s}ek and Hong Yao for various helpful discussions. J.-L.Z. and D.-X.Y. are supported by Grants No. NKRDPC-2017YFA0206203, No. NSFC-11574404, No.NSFG-2015A030313176, the National Supercomputer Center in Guangzhou, and the Leading Talent Program of Guangdong Special Projects. J.-L.Z. is grateful for the hospitality of the Institute for Advanced Study at Tsinghua University and the Kavli Institute for Theoretical Sciences at the University of Chinese Academy of Sciences. W.H. acknowledges financial support from the C.N. Yang Junior Fellowship at Tsinghua University.


\begin{thebibliography}{99}
\bibitem{footnote3} The expressions for $k_{i\mu}^{j\nu}$ in (\ref{eq:kEven}) and (\ref{eq:kOdd}) also involve an integration over the $z$-component fermi momentum $k_{fz}$ for 3D models. It can be checked, separately for the even- and odd-parity pairings, that for the cases where the TRSB pairing simultaneously breaks the two mirror symmetries in question, the integrant is not odd in $k_{fz}$ and hence does not vanish under $k_{fz}$-integration.
\bibitem{footnote2} Only terms with first-order gradient $\partial_y \phi_i$ need to be considered for analyzing the current along $y$. This can be understood as follows. Through a standard minimal coupling to a $U(1)$ gauge field in the free energy, $\partial_\mu \ra \partial_\mu -i a_\mu$, the $y$-current can be expressed as $j_y \propto \left(\frac{\partial f}{\partial a_y} \right)_{\bs a \ra 0}$. Hence higher order $y$-gradients or higher powers of $\partial_y \phi_i$ must have vanishing contribution due to the translation invariance in $y$-direction.
\end{thebibliography}
\end{document}